\documentclass[showpacs,amsmath,amssymb,aps,showkeys,floatfix,prd,a4paper,nofootinbib]{revtex4}

\usepackage[dvips]{graphicx}
\usepackage{dcolumn}
\usepackage{bm}
\usepackage{epsfig}
\usepackage{amsfonts}
\usepackage{amssymb,amscd}
\usepackage{subfigure}

\usepackage{graphicx}

\usepackage[normalem]{ulem}  
\usepackage[dvips]{color} 

\renewcommand\sout{\bgroup \color{red} \ULdepth=-.5ex \ULset}

\def\beq{\begin{equation}}
\def\enq{\end{equation}}

\begin{document}

\preprint{}

\title{$f_0(980)$ production in  $D_s^+ \rightarrow \pi^+ \,  \pi^+ \, \pi^-$ and  
$D_s^+ \rightarrow \pi^+ \,  K^+ \, K^-$  decays}

\author{J.M.~Dias} 
\email{jdias@if.usp.br}
\affiliation{Instituto de F\'{\i}sica, Universidade de S\~{a}o Paulo,
  C.P. 66318, 05389-970 S\~{a}o Paulo, SP, Brazil}

\author{F.~S.~Navarra} 
\email{navarra@if.usp.br}
\affiliation{Instituto de F\'{\i}sica, Universidade de S\~{a}o Paulo,
  C.P. 66318, 05389-970 S\~{a}o Paulo, SP, Brazil}

\author{M.~Nielsen}
\email{mnielsen@if.usp.br}
\affiliation{Instituto de F\'{\i}sica, Universidade de S\~{a}o Paulo,
  C.P. 66318, 05389-970 S\~{a}o Paulo, SP, Brazil}

\author{E.~Oset}
\email{oset@ific.uv.es}
\affiliation{ Departamento de F\'{\i}sica Te\'orica and IFIC, Centro
  Mixto Universidad de Valencia-CSIC, Institutos de Investigaci\'on de
  Paterna, Aptdo. 22085, 46071 Valencia, Spain }

\date{\today}

\begin{abstract}

We study the $D_s^+ \rightarrow \pi^+ \,  \pi^+ \, \pi^-$  and  $D_s^+ \rightarrow \pi^+ \,  
K^+ \, K^-$ decays adopting a mechanism in which the 
$D_s^+$ meson decays weakly into a $\pi^+$ and a $q\bar{q}$ component, which hadronizes into 
two pseudoscalar mesons. The final state interaction between these two pseudoscalar mesons 
is taken into account by using the Chiral Unitary approach in coupled channels, which gives 
rise to the $f_0(980)$ resonance. Hence, we obtain the invariant mass distributions of the 
pairs $\pi^+ \pi^-$ and $K^+ K^-$ after the decay of that resonance and compare our theoretical 
amplitudes with those available from the experimental data. Our results are in a fair 
agreement with the shape of these data, within large experimental uncertainty, and 
a $f_0(980)$ signal is seen in both the $\pi^+\pi^-$ and $K^+K^-$ distributions. Predictions 
for the relative size of $\pi^+\pi^-$ and $K^+K^-$ distributions are made.

\end{abstract}

\pacs{%
}
\maketitle
\section{Introduction}

The analysis of heavy meson weak decays measured in B-factories and at the LHC has been very 
important  for the study of new hadronic states and ultimately for the understanding of hadron 
dynamics. In these reactions, the weak decay leads to hadronic states, 
in general composed by two or three hadrons, which undergo ``final state interactions'' (FSI), 
through which they form the final particles. The FSI is very complex and can influence 
all the conclusions concerning new states and even provide the strength of CP violation 
\cite{Nogueira:2015tsa}. In this work 
we consider the $D_s^+ \rightarrow \pi^+ \,  \pi^+ \, \pi^-$ and  $D_s^+ \rightarrow \pi^+ \,  
K^+ \, K^-$ decays and we study the effect of FSI on the  measured invariant mass spectra. 
These decays have been studied by several experimental groups 
\cite{Onyisi:2013bjt,Zupanc:2013byn,Alexander:2009ux,Alexander:2008aa,
Frabetti:1997sx,delAmoSanchez:2010yp,Aubert:2008ao,Aitala:2000xu} 
and they have been considered  excellent tools to study FSI. Their distinctive feature is the 
fact that they are Cabibbo favored. From the experimental data we know the branching fractions 
\cite{Agashe:2014kda}: 
\beq
\frac{\Gamma(D_s^+ \rightarrow \pi^+ \,  \pi^+ \, \pi^-)}{\Gamma_{\mbox{total}}} = (1.09 
\pm 0.05) \times 10^{-2}\, ;
\label{bra3p}
\enq
\beq
\frac{\Gamma(D_s^+ \rightarrow \pi^+ \,  K^+ \, K^-)}{\Gamma_{\mbox{total}}} = (5.39 \pm 0.21) 
\times 10^{-2}\, .
\label{bra3k}
\enq
The corresponding ratio $ \Gamma(D_s^+ \rightarrow \pi^+ \,  \pi^+ \, \pi^-)) / (\Gamma(D_s^+ 
\rightarrow \pi^+ \,  K^+ \, K^-) \simeq 0.2$ is  in agreement with the value $0.265 \pm 0.041 
\pm 0.031$ found in a previous estimate \cite{Frabetti:1997sx}. 

\begin{figure}[ht!]
\includegraphics[scale=0.30]{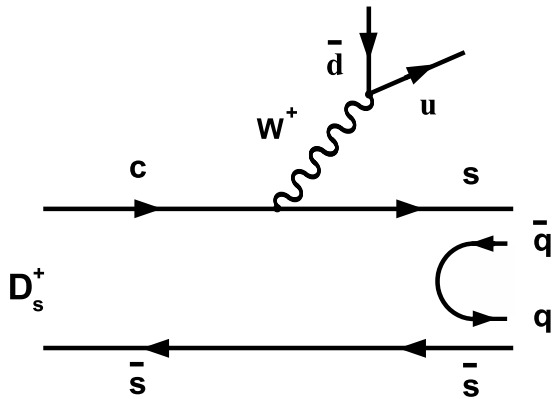} 
\vskip0.5cm
\caption{Schematic representation of the hadronization of the $s\bar{s}$ pair in the 
Cabibbo favored $D_s^+$ weak decay, with the external $\pi^+$ emission. The inserted $\bar{q}q$
pair represents the isoscalar combination $\bar{u}u+\bar{d}d+\bar{s}s$.}
\label{fig1}
\end{figure}

While the differences between these numbers require some quantitive analysis, the qualitative 
relation between these decay rates can be easily understood when we look at the 
Cabibbo favoured decay diagram in Fig.~\ref{fig1}, which is also helicity and color favoured, 
where the $\bar{d}u$ makes up a $\pi^+$ and one has an extra $s\bar{s}$ pair. The final 
$s\bar{s}$ pair hadronizes by creating extra $\bar{q}q$ pairs, which lead to $K\bar{K}$ or 
$\eta \eta$ but not $\pi\pi$. The final state $\pi^+ \,  K^+ \, K^-$ can be produced directly and 
through rescattering  ($K^0 \,  \bar{K}^0 \mbox{ or }\eta\eta\rightarrow K^+ \, K^-$). In contrast, 
the final state $\pi^+ \,  \pi^+ \, \pi^-$ can only be produced  through rescattering ($K^+ \, 
 K^-\mbox{ or }K^0 \,  \bar{K}^0 \mbox{ or }\eta\eta \rightarrow   \pi^+ \, \pi^-$). 

Since the original $s\bar{s}$ pair produced in the Cabibbo favoured $D_s^+$ weak decay,
shown in Fig.~\ref{fig1}, has isospin zero, all the hadrons produced in the hadronization 
process, like the $K\bar{K}$ or $\pi\pi$ final states, have also isospin zero. This means that 
only isospin zero resonances, like $f_0$, can contribute. In the case of the  $K^+K^-$ final state, 
one can also have the contribution of the $\phi$ meson ($K^+K^-$ in $P$ wave). However, the $\rho$ 
meson will not appear in the $D_s^+ \rightarrow \pi^+ \,  \pi^+ \, \pi^-$ decay since the $\rho$ has isospin 1. 
This is of course for the dominant mechanism chosen, but one could expect a small contribution 
from subleading terms. In this work we shall study the processes depicted in Fig.~\ref{fig1}, looking for the 
$f_0(980)$ signal in the spectra of the invariant masses $m_{\pi^+ \pi^-}$ and $m_{K^+ K^-}$.

\section{Formalism}

In order to produce a pair of mesons, the $s\bar{s}$ pair shown in Fig.~\ref{fig1} has to 
hadronize into two mesons. To do that, an extra $\bar q q$ pair with the quantum numbers of the 
vacuum, $\bar{u} u  + \bar{d} d + \bar{s} s$, is added to the already existing quark pair. 
In order to find out the meson-meson components in the hadronized $s\bar{s}$ pair we
define the $q \bar q$ matrix $M$ \cite{alberzou}:
\begin{equation}
\label{eq:1}
M=\left(
           \begin{array}{ccc}
             u\bar u & u\bar d  & u\bar s  \\
             d\bar u & d\bar d  & d\bar s  \\
             s\bar u & s\bar d  & s\bar s  
           \end{array}
         \right) \, ,
\end{equation}
which has the property
\begin{equation}
\label{eq:2}
M \cdot M = M \times ( \bar u u + \bar d d + \bar s s ) .
\end{equation}
The next step consists in writing the matrix $M$ in terms of mesons. Using the standard 
$\eta-\eta^\prime$ mixing \cite{bramon}, the matrix $M$ corresponds to
 \cite{goz}
\begin{equation}\label{eq:3}
\phi=\left(
  \begin{array}{ccc}
    \frac{1}{\sqrt{2}} \pi^0 + \frac{1}{\sqrt{3}} \eta + 
    \frac{1}{\sqrt{6}} \eta ^{\prime}
    & \pi^+  & K^+ \\
    \pi^- &  - \frac{1}{\sqrt{2}} \pi^0 + \frac{1}{\sqrt{3}} \eta + \frac{1}{\sqrt{6}} 
\eta ^{\prime} & K^0 \\
    K^-       & \bar K^0 &  - \frac{1}{\sqrt{3}} \eta + \sqrt{\frac{2}{3}} \eta ^{\prime}  
           \end{array}
         \right).
\end{equation}
Therefore, in terms of two pseudoscalars we have the correspondence:
\begin{equation}
\label{eq:4}
s\bar s  \,  (\bar{u} u  + \bar{d} d + \bar{s} s  )  \equiv  \left( \phi \cdot \phi 
\right)_{33} = K^- K^+ + \bar{K}^0 K^0 + \frac{1}{3} \eta\eta \, ,   
\end{equation}
where we have neglected the $\eta^\prime$ contribution since the mass of $\eta^\prime$ is too
large to be relevant here. These are the states which are produced 
in the first step, prior to FSI. Once a pair of mesons is created they start to
interact and the final $K^+K^-$ or $\pi^+\pi^-$ mesons can be formed as a result of complex
two-body interactions with coupled channels described by the Bethe-Salpeter equation. First 
steps in this direction were given in \cite{gamma} in the $\gamma \gamma \to \mbox{meson-meson}$ 
reaction, proving the accuracy of the method.

In the decay represented in Fig. \ref{fig1} the  $\pi^+$ is treated as a spectator and the 
$s - \bar{s}$ pair may hadronize into $K^+ K^-$, as shown above and, after rescattering, it 
can produce $\pi^+ \pi^-$ and also $K^+ K^-$. The $\pi^+$ that we consider as a spectator 
can also interact  with the $\pi^-$ of the $\pi^+\pi^-$ pair. Yet, investigation of the Dalitz plot 
indicates that the strength of this interaction is shared in a wide region between $530$ MeV 
and $1700$ MeV and thus its contribution in the narrow region of the $f_0(980)$ of the other 
pair is negligible.

The $D_s^+$ decay width into a $\pi^+$ 
and two mesons will be labelled $\Gamma_{P^+P^-}$, where $P^+P^-$ refers to the two
pseudoscalar final mesons: $K^+ K^-$ or $\pi^+ \pi^-$. The differential decay 
width, as a function of the invariant mass of the pair $P^+P^-$ is then given by:
\begin{eqnarray}
\frac{d\Gamma_{P^+P^-}}{dM_{inv}}=\frac{1}{(2\pi)^3}\frac{p_{\pi}\tilde{p}_P}{4M^2_{D_s}}
|T_{P^+P^-} |^2\, ,
\label{decay1}
\end{eqnarray}
where
\begin{equation}
p_{\pi}=\frac{\lambda^{1/2}(M^2_{D_s},m^2_{\pi},M^2_{inv})}{2M_{D_s}} \, ,
\end{equation}
\begin{eqnarray}
\tilde{p}_P=\frac{\lambda^{1/2}(M^2_{inv},m^2_{P^+},m^2_{P^-})}{2M_{inv}}\, ,
\end{eqnarray}
and 
\begin{equation}
\lambda(x^2,y^2,z^2)=x^4+y^4+z^4-2x^2y^2-2x^2z^2-2y^2z^2\, .
\end{equation}
In the above formula $m_{P^+} = m_{K^+}$ or $m_{\pi^+}$ and $m_{P^-} = m_{K^-}$ or 
$m_{\pi^-}$.  The amplitudes in Eq. (\ref{decay1}) are given by 
\begin{eqnarray}
T_{K^+K^-}=V_0~(1+G_{K^+ K^-}\, t_{K^+ K^- \to K^+ K^-} +
G_{K^0 \bar K^0}\, t_{K^0 \bar K^0 \to K^+ K^-}  +
\frac{2}{3}\,\frac{1}{2}\,G_{\eta \eta}\, \tilde{t}_{\eta \eta \to K^+K^-} )\, ,
\label{tkk}
\end{eqnarray}
with $\tilde{t}_{\eta \eta \to K^+K^-}=\sqrt{2}t_{\eta \eta \to K^+K^-}$ and
\begin{eqnarray}
T_{\pi^+\pi^-}=V_0~(G_{K^+ K^-}\, t_{K^+ K^- \to \pi^+ \pi^-} +
G_{K^0 \bar K^0}\, t_{K^0 \bar K^0 \to \pi^+ \pi^-}  +
\frac{2}{3}\,\frac{1}{2}\,G_{\eta \eta}\, \tilde{t}_{\eta \eta \to \pi^+\pi^-} )\, ,
\label{tpipi}
\end{eqnarray}
with $\tilde{t}_{\eta \eta \to \pi^+\pi^-}=\sqrt{2}t_{\eta \eta \to \pi^+\pi^-}$. 
The function $G_{l}$ is the loop function given by:
\begin{equation}
G_{l} (s) = i \int\frac{d^{4}q}{(2\pi)^{4}}\frac{1}{(p-q)^{2}-m^2_{1}+i\varepsilon}\,\frac{1}
{q^{2}-m^2_{2}+i\varepsilon},
\label{eq:G}
\end{equation}
with $p$ being the total four-momentum of the $P^+P^-$ system 
and, hence, the Mandelstam invariant $s$ is $s=p^2=M^2_{inv}$. 
The masses $m_{1}$ and $m_{2}$ are the masses of the mesons 
in the loop for the {\it l}-channel. The factors $2$ and $1/2$ in Eqs.~(\ref{tkk}) and (\ref{tpipi}) 
come from the two combinations to create the $\eta\eta$ state from two $\eta$ fields and the reduction 
of $1/2$ from the loop, all that due to the identity of the two $\eta$ particles. The factor $\sqrt{2}$ relating 
$\tilde{t}$ and $t$ for the $\eta\eta$ channel has the same root in the identity of these two 
particles because for convenience, in the chiral unitary approach the amplitudes $t$ are 
evaluated with the unitary normalization, in this case $|\eta\eta \rangle /\sqrt{2}$ for the $\eta\eta$ state.

The method used here to hadronize the $s \bar{s}$ component and implement 
final state interaction of the resulting meson pair has an early precedent in the 
study of the $J/\psi \to \phi \pi \pi~(K \bar{K})$ decays in \cite{oller}, where 
a relationship between the $s\bar{s}$ and nonstrange form factors and the meson-meson 
interaction was stablished. A different reformulation of the problem, closer to the one 
followed here, is given in \cite{chiang,timo}.

In our calculations the integral 
on $q^0$ in Eq.~(\ref{eq:G}) is performed exactly analytically and a cut-off, $|\vec{q}_{max}| 
= 600~ \rm{MeV}/c$, is introduced in the integral on $\vec{q}$.
The elements of the scattering matrix $t_{i\to j}$ are the solutions 
of the Bethe-Salpeter equation. Namely, we obtain these elements 
by solving a coupled-channel scattering equation in an algebraic form
\begin{equation}
t_{i\to j} (s) = V_{i j} (s) + \sum^{5} _{l=1} V_{i l} (s) G_{l} (s) t_{l\to j} (s) ,
\label{eq:BSEq}
\end{equation}
where each value assumed by the $i$, $j$, and $l$ indices in 
the range from 1 to 5 indicates the channels: 1 for $\pi^+ \pi^-$, 2 
for $\pi^0 \pi^0$, 3 for $K^+ K^-$, 4 for $K^0 \bar K^0$ and 5 for 
$\eta \eta$. $V$ is the interaction kernel which corresponds to 
the tree-level transition amplitudes obtained from phenomenological 
Lagrangians developed in Ref.~\cite{Oller:1997ti}, complemented 
with the inclusion of the matrix elements for the $\eta \eta$ channels 
given in \cite{Gamermann:2006nm}. This cut off of $600$ $\mbox{MeV/c}$, 
different from  the one used in \cite{Oller:1997ti}, is needed to reproduce 
experimental amplitudes when the $\eta\eta$ channel is introduced 
explicitly \cite{liang,dai}. We have
\begin{align}
&V_{11}=-\frac{1}{2f^2}s,~&&V_{12}=-\frac{1}{\sqrt{2}f^2}(s-m_{\pi}^2),~&&V_{13}=-\frac{1}{4f^2}s,\nonumber \\
&V_{14}=-\frac{1}{4f^2}s,~&&V_{15}=-\frac{1}{3\sqrt{2}f^2}m_{\pi}^2,~&&V_{22}=-\frac{1}{2f^2}m_{\pi}^2,\nonumber \\
&V_{23}=-\frac{1}{4\sqrt{2}f^2}s,~&&V_{24}=-\frac{1}{4\sqrt{2}f^2}s,~&&V_{25}=-\frac{1}{6f^2}m_{\pi}^2,\nonumber \\
&V_{33}=-\frac{1}{2f^2}s,~&&V_{34}=-\frac{1}{4f^2}s,~&&V_{35}=-\frac{1}{12\sqrt{2}f^2}(9s-6m_{\eta}^2-2m_{\pi}^2),\nonumber \\
&V_{44}=-\frac{1}{2f^2}s,~&&V_{45}=-\frac{1}{12\sqrt{2}f^2}(9s-6m_{\eta}^2-2m_{\pi}^2),~&&V_{55}=-\frac{1}{18f^2}(16m_{K}^2-7m_{\pi}^2), 
\label{eq:Vkernel}
\end{align}
where $f$ represents the pion decay constant, $f=f_{\pi}=93$ MeV, and $m_{\pi}$, 
$m_{K}$, and $m_{\eta}$ are the averaged masses of pion, 
kaon, and $\eta$ mesons, respectively.

\begin{figure}[htb!] 
\includegraphics[scale=0.9]{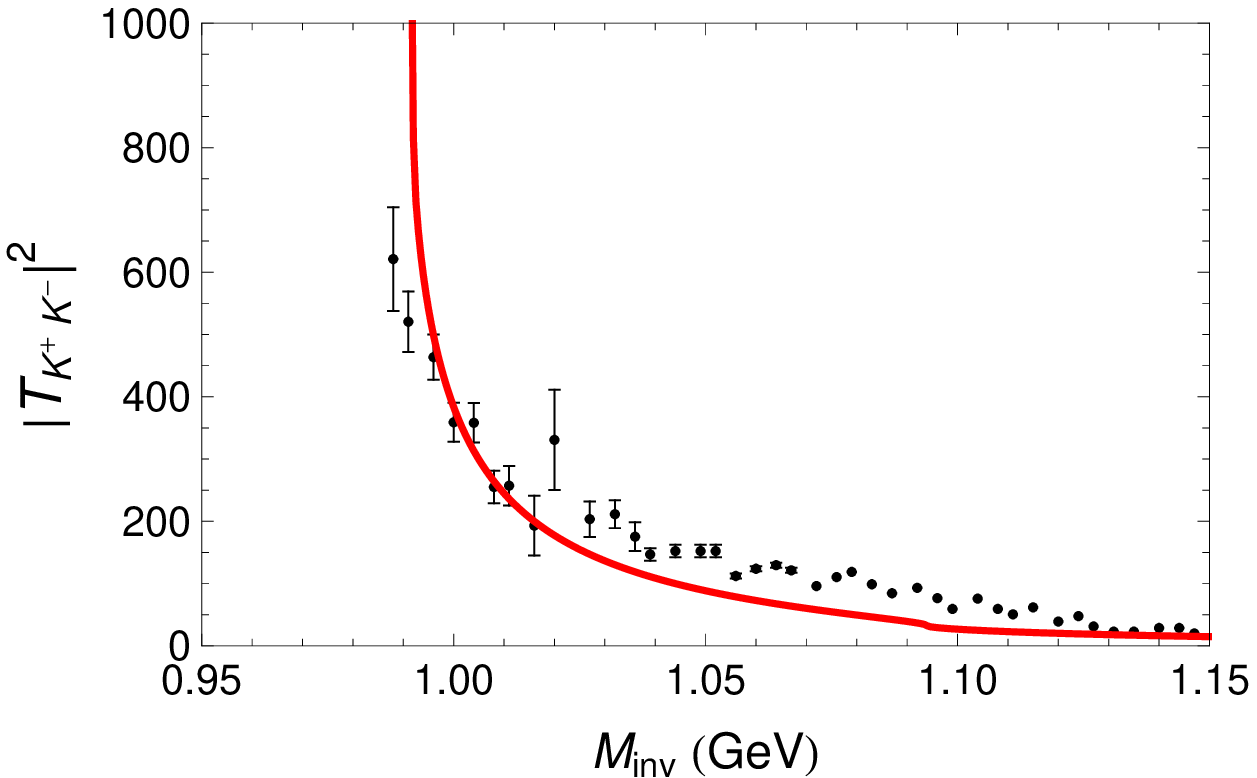}  
\vskip0.5cm 
\caption{$|T_{K^+K^-}|^2$  as a function of the $K^+K^-$ invariant mass as obtained from Eq.~(\ref{tkk}) (solid 
line). The experimental data are taken from 
\cite{delAmoSanchez:2010yp}.} 
\label{a2kk} 
\end{figure} 

The large overlap of the $f_0(980)$, and small one for the $f_0(500)$, with the 
$s\bar{s}$ components was emphasized in \cite{scadron}, where using the linear $\sigma$ model 
and a mixing of strange and non strange $q\bar{q}$ components, a qualitative description of the 
$\phi \to \pi^0 \pi^0 \gamma$, $f_0(980) \to \gamma \gamma$, $J/\psi \to \omega \pi \pi$ was given. 
Posterior work using chiral Lagrangians for the meson meson interactions, together with unitarization 
in coupled channels, has produced precise quantitative descriptions of these and many other reactions. 
The $\phi \to \pi^0 \pi^0 \gamma$ reaction was studied in \cite{marco,palomar,hanhart}. The 
$J/\psi \to \omega \pi \pi$ and $J/\psi \to \phi \pi \pi$ reactions were studied in 
\cite{oller,chiang,timo}, obtaining a quantitative description of the spectra, and explaining 
why the $f_0(500)$ is seen in the first reaction and the $f_0(980)$ in the second 
one. The $f_0(500)$ and $f_0(980)$ coupling to $\gamma \gamma$ was studied in 
\cite{Oller:1997ti,gamma,roca}. A review on these and related issues using the chiral unitary approach is 
given in \cite{ramosoller}. In the present work we have used the chiral unitary approach of \cite{Oller:1997ti}, 
which has proved to be a precise tool to account for strong interactions at low energies \cite{ramosoller}.

\section{Results}

The numerical results for the amplitude squared $|T_{K^+K^-}|^2$ 
as a function of the $K^+K^-$ invariant mass, as obtained from 
Eq.~(\ref{tkk}), are shown in Fig.~\ref{a2kk}. In this figure we also 
show the experimental data for the $S$-wave contribution for the 
$K\bar{K}$ mass distribution extracted from Ref.~\cite{delAmoSanchez:2010yp}. 
We adjust the $V_0$ parameter in Eq.~(\ref{tkk}) to approximately fit the data. 
The theoretical curve represents 
essentially $|t^{I=0}_{K\bar{K}\to K\bar{K}}|^2$, which is dominated by 
the $f_0(980)$ pole in that region. Indeed, since (recall that $|K^-\rangle=-|1/2\, -1/2\rangle$)
\begin{eqnarray}
|K\bar{K}, I=0, I_3=0\rangle = -\frac{1}{\sqrt{2}}(K^+ K^- + K^0 \bar{K}^0)\, ,\nonumber \\
|K^+ K^-\rangle = -\frac{1}{\sqrt{2}}(|K\bar{K}, I=1\rangle + |K\bar{K}, I=0\rangle)\, ,
\end{eqnarray}
then 
\begin{equation}
t_{K^+K^-\to K^+ K^-}+t_{K^0\bar{K}^0\to K^+ K^-}=t^{I=0}_{K\bar{K}\to K\bar{K}}\, ,
\end{equation}
and from the Bethe-Salpeter equation, ignoring $\eta\eta \to K^+K^-$, we have
\begin{equation}
\frac{T_{K^+K^-}}{V_0}\equiv 1+G \, t^{I=0}_{K\bar{K}\to K\bar{K}}\simeq
\frac{V^{I=0}+V^{I=0}\,G\, t^{I=0}}{V^{I=0}}=\frac{t^{I=0}_{K\bar{K}\to 
K\bar{K}}}{V^{I=0}}\, ,
\end{equation}
where $V^{I=0}$ is the $K\bar{K}\to K\bar{K}$ potential in $I=0$. The sign $\simeq$ is used 
because we also ignore the $\pi \pi$ channel in that equation, which plays a minor role 
around the $f_0(980)$ region. Thus, Eq.~(\ref{tkk}) is roughly 
proportional  to $T_{K^+K^-}$, which reflects the $f_0(980)$ resonance in this region.

We have chosen to reproduce the data around $1$ GeV and the agreement looks fair above this 
energy, but clear discrepancies are seen for smaller values of $M_{inv}$. The discrepancies 
with the data are unavoidable because in \cite{delAmoSanchez:2010yp} a mass of $922$ MeV and width of $240$ MeV 
were obtained for the $f_0(980)$, while our calculations provide results in good agreement with the 
PDG average. The PDG results are $M=(980 \pm 20)$ MeV, $\Gamma=(40  - 100)$ MeV. From the 
$B_s \to J/\psi \pi \pi$ reaction one obtains similar results $M=(972\pm 20)$ MeV, $\Gamma 
\approx 50$ MeV. The clear discrepancies of \cite{delAmoSanchez:2010yp} with the standard results 
should be enough motivations to look again at this reaction with more detail.

\begin{figure}[htb!] 
\includegraphics[scale=0.9]{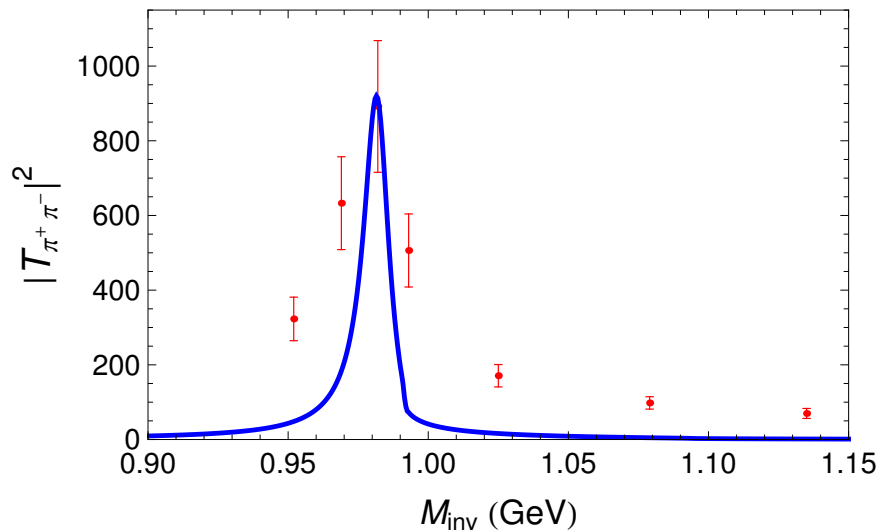}
\vskip0.5cm 
\caption{$|T_{\pi^+\pi^-}|^2$  as a function of the  $\pi^+\pi^-$ invariant mass  
as obtained from Eq.~(\ref{tpipi}) (solid line). The experimental data 
are taken from \cite{Aubert:2008ao}.}
\label{a2pipi} 
\end{figure} 

In Fig.~\ref{a2pipi} we show $|T_{\pi^+\pi^-}|^2$ as a function 
of the invariant mass of the pair of pions $\pi^+\pi^-$ obtained 
from Eq.~(\ref{tpipi}), and the experimental data for the $S$-wave 
contribution for the $\pi\pi$ mass distribution extracted from 
\cite{Aubert:2008ao}. The normalization of the $K^+K^-$ production rates 
divided by the phase space of \cite{delAmoSanchez:2010yp} and the normalization 
of the $\pi^+\pi^-$ production divided by the phase space of \cite{Aubert:2008ao} 
are not the same. In Ref.~\cite{delAmoSanchez:2010yp} the distributions 
are superposed (see Fig.~6 of that reference) to show that their ``profile'' around the 
$f_0(980)$ is the same. We can normalize the value of $V_0$ to these 
$\pi^+\pi^-$ data. At the $f_0(980)$ peak position 
our theoretical curve agrees with the data, by construction, but for lower and 
higher values of the $\pi^+\pi^-$ invariant mass the experimental 
distribution is broader than the theoretical calculation. One reason 
for that could be the fact that in the experimental data they found a 
$S$-wave contribution from the $f_0(1370)$ and 
$f_0(1500)$ resonances, which are not included in this calculation. 

\begin{figure}[htb!] 
\includegraphics[scale=0.85]{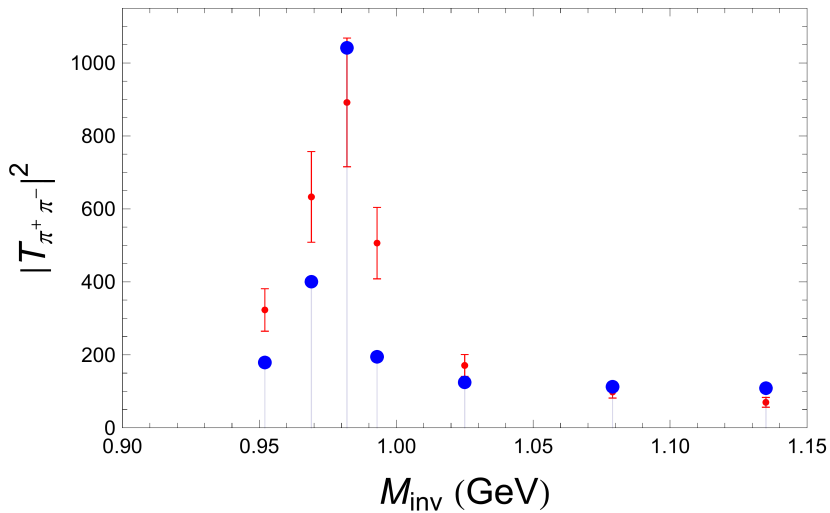}
\vskip0.5cm 
\caption{$|T_{\pi^+\pi^-}|^2$  as a function of the $\pi^+\pi^-$ invariant mass. Here, the theoretical 
results (thick circles without error bars) are folded in order to have the same size of the 
experimental bins, which is $25$ MeV. The experimental data are taken from \cite{Aubert:2008ao}.}
\label{t2pipibin} 
\end{figure}

It is worth emphasizing that our calculations do not take into account sources of 
background, which would come, as we discussed earlier, from the consideration 
of the interaction of the spectator pion with the other pions.
Moreover, we should note that in 
Ref.~\cite{Aubert:2008ao} the authors have bins of about $15$ MeV or more, which 
are used to construct the few $\pi^+\pi^-$ experimental points of the mass 
distribution. Thus, in order to have a better comparison with 
data, we do two things. First, we integrate our mass distribution over the same bins as 
experiment, dividing by the size of the bins. Second, we add a background to our results. 
This background is chosen constant in $M_{inv}$, such as to get a fair reproduction of the 
last three experimental points. The result is shown in Fig.~\ref{t2pipibin}. Now, 
the agreement with the data looks better than in Fig.~\ref{a2pipi}, but still our distribution 
seems a bit narrower than the experimental one. It is also worth mentioning that for this latter 
observable, the theoretical work of \cite{liang} also misses some strength on the 
sides of the $f_0(980)$ resonance with respect to the experimental data of 
Ref.~\cite{lhcb}. One approach based on the use of the $s\bar{s}$ pion 
form factor, obtained with an Omnes representation constructed from experimental pion-pion phase 
shifts, fills up this region \cite{Daub:2015xja}. An alternative approach that could be 
tested in these reactions is the one used in Ref.~\cite{weiwang2} using light cone 
sum rules to evaluate form factors, together with unitarization of the final meson pairs. 
It is also worth mentioning that if we extrapolate 
the $\pi^+\pi^-$ distribution to lower invariant masses we do not find a trace of the $f_0(500)$. This 
feature is also noted in Ref.~\cite{Aubert:2008ao} and it was also the case in the 
$B_s^0\to J/\psi \pi^+\pi^-$ experiment \cite{lhcb}, as well as in the theoretical 
descriptions in Ref.~\cite{liang,Daub:2015xja,weiwang}.

\begin{figure}[htb!] 
\vskip0.5cm
\includegraphics[scale=1.2]{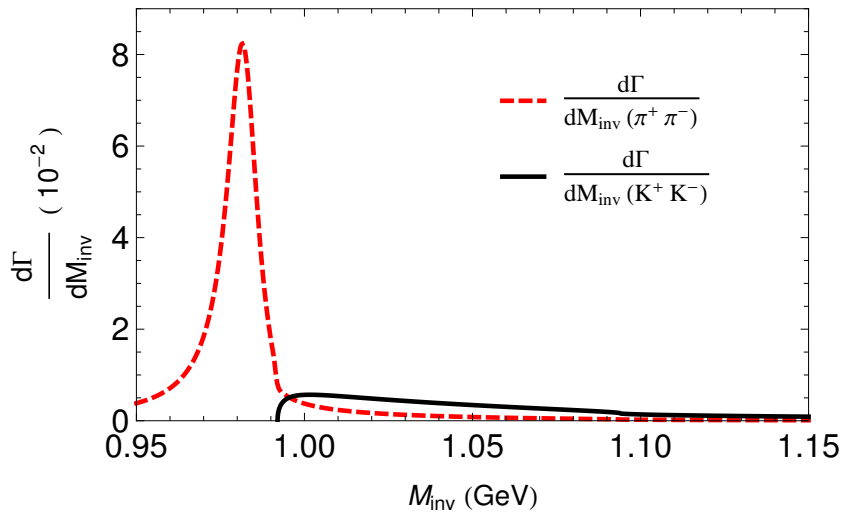}
\vskip0.5cm 
\caption{The $\pi^+\pi^-$ and $K^+K^-$ invariant mass distributions 
for the $D^+_s\to \pi^+\pi^-\pi^+$ (dashed line) and $D^+_s\to \pi^+K^-K^+$ 
(solid line) with arbitrary normalization, respectively.} 
\label{dkdpi} 
\end{figure} 

So far we have discussed only  the shapes of the  $|T_{\pi^+\pi^-}|^2$ and $|T_{K^+K^-}|^2$ 
amplitudes and  now we wish to make predictions for the relative strength of the rates 
of the two reactions. We  can use Eq.~(\ref{decay1}) in order to predict the 
$\pi^+\pi^-$ and $K^+K^-$ distributions, as illustrated in 
Fig.~\ref{dkdpi}. According to this figure, we see the $f_0(980)$ 
signal in the spectra of the invariant mass $m_{\pi^+\pi^-}$, 
as indicated by the dashed curve. On the other hand, the $K^+K^-$ 
distribution gets strength from the underlying $f_0(980)$ resonance 
close to the $K^+K^-$ threshold. It would be most useful to determine experimentally 
the strength of these two distributions to compare with these predictions, which, up to 
a global common normalization factor, are predictions of the Chiral Unitary approach 
with no free parameters.

One should stress once more that the predictions are limited to the region close to 
$f_0(980)$. In principle one should study dynamics involving three meson interactions 
\cite{alberto}, but, as discussed earlier, the wide range of invariant masses of the spectator 
$\pi^+$ with any of those producing the $f_0(980)$ dilutes its contribution into a 
background. As for the $K\bar{K}$ distribution in Fig.~\ref{dkdpi} one should also note 
that, if one goes to higher invariant masses, the method used here would have to be 
complemented with extra channels that are for instance discussed in \cite{ollerplus,geng}.

The results of Fig.~\ref{dkdpi} might look in conflict with the ratio obtained from the data in 
Eqs.~(\ref{bra3p}) and (\ref{bra3k}). We mentioned in the introduction that from these one finds the 
$\pi^+ K^+ K^-$ rate to be about five times larger than the one of $\pi^+ \pi^+ \pi^-$. The results of 
Fig.~\ref{t2pipibin} in the range of $M_{inv}$ of the figure are opposite and the $\pi^+ \pi^+ \pi^-$ 
strength is bigger than the one of $K^+K^-$. The discrepancy is only apparent because the rates 
of Eq.~(\ref{bra3p}) and (\ref{bra3k}) extend to all the range of invariant masses and for any possible 
partial wave. We only consider $s$-wave, which can be disentangled in an experimental analysis. For 
instance, the $D_s^+ \to \pi^+ K^+ K^-$ decay gets a large contribution, from 
$D_s^+\to \pi^+ \phi$ ($\phi \to K^+ K^-$), with a branching ratio of $2.27\times 10^-2$ 
\cite{Agashe:2014kda}, which we do not consider, and there are contribution from higher mass 
resonances that couple to $K\bar{K}$. Our predictions are limited to low values of $M_{inv}$ 
close to the $K\bar{K}$ threshold, and exclude the $P$-wave $\phi$ production followed 
by $\phi \to K^+ K^-$.

\section{A discussion on the Tetraquark picture and the present reaction}

Concerning the $f_0(980)$ and other light scalar mesons, $f_0(500)$, $a_0(980)$, $\kappa(800)$, there is 
much discussion about their nature as $q\bar{q}$, tetraquark, molecules, dynamically generated states, etc 
\cite{klempt}. What seems to have reached the consensus of the scientific community is that they are not ordinary, 
$q\bar{q}$, mesons (see extensive information on the subject in the report \cite{sigma}). There is more discussion 
on whether they are tetraquarks or they appear dynamically generated from the meson meson interaction, the picture 
we have adopted here, and which we implement using the chiral unitary approach.

The tetraquark picture for mesons developed in \cite{jaffe} has been extensively used in the literature concerning the 
scalar mesons \cite{schechter,polosa,narison,achasov,stone}. The most common configuration is given for the 
$f_0(500)$ and $f_0(980)$ by
\begin{equation}\label{f0s}
f_0(500)=[ud][\bar{u}\bar{d}] \, \, , \quad
f_0(980)=\frac{1}{\sqrt{2}}([su][\bar{s}\bar{u}]+[sd][\bar{s}\bar{d}]) \, ,
\end{equation}
by means of which one finds a qualitative description of the masses of these mesons. There are also problems since 
the $f_0(980)$ does not couple to $\pi\pi$ in the picture of Eq.~\eqref{f0s} and the coupling $f_0\pi\pi$ is too small 
compared with experiment even if some configuration mixing is considered \cite{polosa}. This means that in those 
pictures one would get a very small $D_s^+\to \pi^+\pi^+\pi^-$ rate compared to $D_s^+\to \pi^+K^+K^-$ with respect 
to our predictions in Fig.~\ref{dkdpi}. In some refinements to the basic model, new elements are introduced to solve 
one or another problem related to phenomenology. In \cite{polosa} the authors include instanton components to fix 
the $\pi\pi$ coupling problem. In \cite{narison} $q\bar{q}$-gluonium components are introduced to address the 
problem of $\pi\pi\to\pi\pi$, $\pi\pi\to K\bar{K}$ and $\gamma\gamma\to \pi\pi$ scattering. In \cite{achasov} in order to 
reproduce the data of the $\phi\to \pi^0\pi^0\gamma$ reaction, the tetraquark picture is also invoked, but the 
$f_0(980)$ is claimed to be largely made of the $[sd][\bar{s}\bar{d}]$ component. Some basic features of spectra 
can be related to the fact that there are four quarks, independent of the particular rearrangements \cite{schechter}. What seems 
to be missing in this approach is a unique picture that describes all processes where these mesons appear, 
instead of invoking different dynamical aspects for each one of them.

In this respect it is interesting to mention that, using the picture of Eq.~\eqref{f0s} for the tetraquarks, it was shown 
in \cite{stone} that it was not possible to reconcile the ratios of decay rates to $f_0(500)$ and $f_0(980)$ seen in the 
$B^0\to J/\psi \pi^+\pi^-$ and $B_s^0\to J/\psi \pi^+\pi^-$ decays, which have a large signal for the $f_0(980)$ in 
$B_s^0\to J/\psi \pi^+\pi^-$ decay and practically no $f_0(500)$, while the reverse situation is found in 
$B^0\to J/\psi \pi^+\pi^-$ \footnote[1]{In \cite{stone} the results were found compatible with a $q\bar{q}$ picture, but the overwhelming evidence 
against it from the discussions in \cite{sigma} do not make this coincidence a case in favor of the $q\bar{q}$ 
picture for the scalars.}.

There is also one feature that cannot escape this discussion. In physical processes involving these resonances one 
looks for $\pi\pi$ or $K\bar{K}$ in the final states. Independently of the dynamics generating the resonances, the 
$\pi\pi$ or $K\bar{K}$ will undergo final state interaction, scattering and making transitions among them, something 
that is not normally accounted for in the tetraquark pictures. Also some reactions have large contributions from tree level 
production of pairs of mesons, which can revert into $K\bar{K}$ or $\pi \pi$ at the end through rescattering, and this 
dynamics escapes the description of the process in terms of tetraquarks alone.

Accepting that some of the dynamics on the tetraquarks models is well founded, our approach is different and does not 
necessarily contradict it. Our approach starts accepting that QCD dynamics at low energies is governed by the effective 
chiral Lagrangians \cite{weinberg}. From these Lagrangians we construct the leading terms of the meson meson interaction 
and then, using a unitary chiral approach in coupled channels, we generate the full meson meson amplitudes. In s-wave 
and $I=0$ these amplitudes contain poles which correspond to the $f_0(500)$ and $f_0(980)$ resonances. In $I=1$ 
the $\eta \pi$, $K\bar{K}$ amplitudes  generate the $a_1(980)$ resonance and the $\pi K$ and $\eta K$ give rise to the 
$\kappa(800)$. All this is obtained with only one parameter which is needed to regularise the loops. Hence, the approach 
contains the scalar mesons and the scattering amplitudes needed to face different problems where the resonances are 
produced. It is most rewarding to see that the problems mentioned above, that required the introduction of different elements 
in the tetraquark pictures, are well described in this unified picture. In this sense, the $\phi\to\pi^0\pi^0\gamma$, 
$\pi^+\pi^-\gamma$, $\pi^0\eta\gamma$ reactions are described within this picture in \cite{marco}. The couplings of 
the $f_0(500)$, $f_0(980)$ to $\pi\pi$, $K\bar{K}$ are obtained in \cite{Oller:1997ti} and \cite{nsd} in agreement with 
phenomenology. The $\gamma\gamma\to \pi\pi$ reaction is also addressed successfully within this picture in \cite{gamma} 
and the puzzle addressed in \cite{stone} concerning the $B^0\to J/\psi\pi\pi$ and $B_s^0\to J/\psi\pi\pi$ reactions 
was properly described in this picture in \cite{liang}. These are only a few examples of cases  where the chiral unitary 
approach proves most suited to describe the physical processes where the scalar resonances are produced. A more complete 
description can be obtained in the review papers \cite{ramosoller} and \cite{review}.

\section{Conclusions}

In this paper we addressed the study of the $D_s^+$ decays into $\pi^+\pi^+\pi^-$ and 
$\pi^+K^+K^-$ mesons. The $\pi^+\pi^-$ and $K^+K^-$ meson 
pairs in the final state were allowed to undergo interactions in coupled channels and lead to 
the $f_0(980)$ resonance production. We adopted a mechanism which involves the $D_s^+$ 
weak decays into a $\pi^+$ and a $q\bar{q}$ component, that is Cabibbo and also color favoured. 
Upon hadronization of the $q\bar{q}$ component into a pair of two pseudoscalar mesons, the final 
state interaction between them is taken into account by using the Chiral Unitary theory where 
$f_0(980)$ emerges as a dynamically generated resonance, which then decays into $\pi^+\pi^-$ 
and also into $K^+K^-$ mesons. In order to do that, we solved the Bethe-Salpeter equation in 
coupled channels. We observe that our curves for the $|T_{\pi^+\pi^-}|^2$ and $|T_{K^+K^-}|^2$ 
amplitudes, obtained as a function of the $\pi^+\pi^-$ and $K^+K^-$ invariant masses, 
respectively, have a shape in fair agreement with the data reported in 
Refs.~\cite{delAmoSanchez:2010yp,Aubert:2008ao}, with the unavoidable discrepancies for 
$|T_{K^+K^-}|^2$ at low masses because of the small mass of the $f_0(980)$ obtained 
in \cite{delAmoSanchez:2010yp} of $922$ MeV. To the best of our knowledge, in the present work  
these data are for the first time addressed from the theoretical point of view. 

We could also determine the shape and 
strength of the $\pi^+\pi^-$ or $K^+K^-$ mass distributions in those two reactions, which, 
up to a common global normalization constant, are a prediction of the theoretical 
approach with no further parameters.

These decays provide an important scenario to test the predictions 
of the chiral unitary theory as well as the nature of the $f_0(980)$ resonance, 
since the latter emerges from this approach after taking into account the interaction between two 
pseudoscalar mesons, which generates dynamically the low lying scalar mesons. So far 
only shapes for these reactions have been established experimentally. The measurement 
of the relative strength of these two mass distributions would be most welcome to contrast 
them with the theoretical predictions.

Another interesting issue would be to study the $\pi^+\pi^0 \eta$ decay mode. 
This would generate the $a_0(980)$ and one could address again the issue 
of the $f_0(980)$ and $a_0(980)$ mixing \cite{f0a0mixing}. This would be obtained in our 
formalism by taking different masses of the charged and neutral kaons in the loop function $G$ for 
$K\bar{K}$, as done in \cite{aceti}.

\begin{acknowledgments}

  We thank Ignacio Bediaga, Alberto Correa dos Reis and A.~M.~Torres for useful
  discussions and comments.  We acknowledge the support by FAPESP, CNPq and by Open
  Partnership Joint Projects of JSPS Bilateral Joint Research
  Projects.  This work is partly supported by the Spanish Ministerio
  de Economia y Competitividad and European FEDER funds under the
  contract number FIS2011-28853-C02-01 and FIS2011-28853-C02-02, and
  the Generalitat Valenciana in the program Prometeo II-2014/068.  

\end{acknowledgments}



\begin{thebibliography}{99}

\bibitem{Nogueira:2015tsa} 
  J.~H.~Alvarenga Nogueira, I.~Bediaga, A.~B.~R.~Cavalcante, T.~Frederico and O.~Lourenço,\\
  Phys.\ Rev.\ D {\bf 92}, 054010 (2015).

\bibitem{Onyisi:2013bjt} 
  P.~U.~E.~Onyisi {\it et al.} [CLEO Collaboration],
  Phys.\ Rev.\ D {\bf 88}, 032009 (2013). 

\bibitem{Zupanc:2013byn} 
  A.~Zupanc {\it et al.} [Belle Collaboration],
  JHEP {\bf 1309}, 139 (2013).

\bibitem{Alexander:2009ux} 
  J.~P.~Alexander {\it et al.} [CLEO Collaboration],
  Phys.\ Rev.\ D {\bf 79}, 052001 (2009).

\bibitem{Alexander:2008aa} 
  J.~P.~Alexander {\it et al.} [CLEO Collaboration],
  Phys.\ Rev.\ Lett.\  {\bf 100}, 161804 (2008). 

\bibitem{Frabetti:1997sx} 
  P.~L.~Frabetti {\it et al.} [E687 Collaboration],
  Phys.\ Lett.\ B {\bf 407}, 79 (1997).

\bibitem{delAmoSanchez:2010yp} 
  P.~del Amo Sanchez {\it et al.} [BaBar Collaboration],
  Phys.\ Rev.\ D {\bf 83}, 052001 (2011); 

\bibitem{Aubert:2008ao}
  B.~Aubert {\it et al.} [BaBar Collaboration],
  Phys.\ Rev.\ D {\bf 79} (2009) 032003.
  
\bibitem{Aitala:2000xu} 
  E.~M.~Aitala {\it et al.} [E791 Collaboration],
  Phys.\ Rev.\ Lett.\  {\bf 86}, 770 (2001).

\bibitem{Agashe:2014kda}   K.~A.~Olive {\it et al.} [Particle Data Group Collaboration],
  Chin.\ Phys.\ C {\bf 38}, 090001 (2014).

\bibitem{alberzou}  A.~Martinez Torres, L.~S.~Geng, 
L.~R.~Dai, B.~X.~Sun, E.~Oset and B.~S.~Zou,
Phys.\ Lett.\ B {\bf 680}, 310 (2009). 
                    
\bibitem{bramon}
  A.~Bramon, A.~Grau and G.~Pancheri,
  Phys.\ Lett.\ B {\bf 283}, 416 (1992).
  
  
\bibitem{goz} D.~Gamermann, E.~Oset, B.S. Zou,  Eur.\ Phys.\ J.\ A {\bf 41}, 85 (2009).

\bibitem{gamma} 
  J.~A.~Oller and E.~Oset,
  Nucl.\ Phys.\ A {\bf 629}, 739 (1998)
  
\bibitem{oller} 
  U.~G.~Meissner and J.~A.~Oller,
  Nucl.\ Phys.\ A {\bf 679}, 671 (2001)
    
\bibitem{chiang} 
  L.~Roca, J.~E.~Palomar, E.~Oset and H.~C.~Chiang,
  Nucl.\ Phys.\ A {\bf 744}, 127 (2004)

\bibitem{timo} 
  T.~A.~Lahde and U.~G.~Meissner,
  Phys.\ Rev.\ D {\bf 74}, 034021 (2006)
  

\bibitem{Oller:1997ti} 
  J.~A.~Oller and E.~Oset,
  Nucl.\ Phys.\ A {\bf 620}, 438 (1997); 
  [Nucl.\ Phys.\ A {\bf 652}, 407 (1999)].
  
\bibitem{Gamermann:2006nm} 
  D.~Gamermann, E.~Oset, D.~Strottman and M.~J.~Vicente Vacas,
  Phys.\ Rev.\ D {\bf 76}, 074016 (2007).
  
\bibitem{dai} 
  J.~J.~Xie, L.~R.~Dai and E.~Oset,
  Phys.\ Lett.\ B {\bf 742}, 363 (2015).  
  
\bibitem{liang} 
  W.~H.~Liang and E.~Oset,
  Phys.\ Lett.\ B {\bf 737}, 70 (2014). 
    
\bibitem{scadron} 
  R.~Delbourgo, D.~s.~Liu and M.~D.~Scadron,
  Phys.\ Lett.\ B {\bf 446}, 332 (1999)
      
\bibitem{marco} 
  E.~Marco, S.~Hirenzaki, E.~Oset and H.~Toki,
  Phys.\ Lett.\ B {\bf 470}, 20 (1999)
  
\bibitem{palomar} 
  J.~E.~Palomar, L.~Roca, E.~Oset and M.~J.~Vicente Vacas,
  Nucl.\ Phys.\ A {\bf 729}, 743 (2003)

\bibitem{hanhart} 
  Y.~S.~Kalashnikova, A.~E.~Kudryavtsev, A.~V.~Nefediev, C.~Hanhart and J.~Haidenbauer,
  Eur.\ Phys.\ J.\ A {\bf 24}, 437 (2005)
  
 
\bibitem{roca} 
  J.~A.~Oller and L.~Roca,
  Eur.\ Phys.\ J.\ A {\bf 37}, 15 (2008)
 
   
\bibitem{ramosoller} 
  J.~A.~Oller, E.~Oset and A.~Ramos,
  Prog.\ Part.\ Nucl.\ Phys.\  {\bf 45}, 157 (2000)
       
\bibitem{lhcb} 
  R.~Aaij {\it et al.} [LHCb Collaboration],
  Phys.\ Rev.\ D {\bf 89}, 092006 (2014). 

\bibitem{Daub:2015xja} 
  J.~T.~Daub, C.~Hanhart and B.~Kubis,
  JHEP {\bf 1602}, 009 (2016)
  
 \bibitem{weiwang2} 
  U.~G.~Meissner and W.~Wang,
  Phys.\ Lett.\ B {\bf 730}, 336 (2014) .
  
\bibitem{weiwang} 
  Y.~J.~Shi and W.~Wang,
  Phys.\ Rev.\ D {\bf 92}, 074038 (2015).
  
\bibitem{alberto}
A.~M.~Torres, K.~P.~Khemchandani, L.~S.~Geng, M. Napsupiale, and E.~Oset, 
Phys.\ Rev.\ {\bf D 78}, 074031 (2008).
  
\bibitem{ollerplus} 
  M.~Albaladejo and J.~A.~Oller,
  Phys.\ Rev.\ Lett.\  {\bf 101}, 252002 (2008)
  
\bibitem{geng}
L.~S.~Geng, E.~Oset, Phys.\ Rev.\ {\bf D 79}, 074009 (2009).    
  
\bibitem{klempt}
E.~Klempt, A.~Zaitsev, 
Phys.\ Rept.\ 454, 1-202 (2007).  

\bibitem{sigma}
J.~R.~Pelaez, arXiv:1510.00653 (2015). To be published in Physics Reports.
  
\bibitem{jaffe}
R.~L.~Jaffe, Phys.\ Rev.\ D {\bf 15}, 267 (1977).

R.~.L~ Jaffe, Phys.\ Rev.\ D {\bf 15}, 281 (1977).  

\bibitem{schechter}
A.~H.~Fariborz, R.~Jora, J.~Schechter, 
Phys.\ Rev.\ D {\bf 79}, 074014 (2009).

\bibitem{polosa}
G.~t' Hooft, G.~Isidori, L.~Maiani, A.~D.~Polosa, V.~Riquer,
Phys.\ Lett.\ B {\bf 662}, 424-430 (2008).

\bibitem{narison}
G.~Mennessier, S.~Narison, and X.~-G.~Wang, 
Phys.\ Lett.\ B {\bf 696} (2011) 40.

\bibitem{achasov}
N.~N.~Achasov, A.~V.~Kiselev, 
Phys.\ Rev.\ D {\bf 83}, 054008.

\bibitem{stone}
S.~Stone, L.~Zhang, 
Phys.\ Rev.\ Lett.\ 111, 062001 (2013).

\bibitem{weinberg}
 S.~Weinberg, 
 Physica A {\bf 96}, 327-340 (1979). 
 
\bibitem{nsd}
J.~A.~Oller, E.~Oset, 
Phys.\ Rev.\ D\ {\bf 60}, 074023 (1999). 

\bibitem{review}
E.~Oset {\it et al}.,
Int.\ J.\ Mod.\ Phys.\ E {\bf 25}, 1630001 (2016).
  
\bibitem{f0a0mixing}
N.~N.~Achasov, S.~A.~Devyanin and G.~N.~Shestakov,
  Phys.\ Lett.\ B {\bf 88}, 367 (1979);

J.~J.~Wu, Q.~Zhao and B.~S.~Zou,
  Phys.\ Rev.\ D {\bf 75}, 114012 (2007);

C.~Hanhart, B.~Kubis and J.~R.~Pelaez,
  Phys.\ Rev.\ D {\bf 76}, 074028 (2007);

 J.~J.~Wu and B.~S.~Zou,
  Phys.\ Rev.\ D {\bf 78}, 074017 (2008);

 M.~Ablikim {\it et al.} [BESIII Collaboration],
  Phys.\ Rev.\ D {\bf 83}, 032003 (2011);

 L.~Roca,
  Phys.\ Rev.\ D {\bf 88}, 014045 (2013);

V.~E.~Tarasov, W.~J.~Briscoe, W.~Gradl, A.~E.~Kudryavtsev and I.~I.~Strakovsky,
  Phys.\ Rev.\ C {\bf 88}, 035207 (2013);
   

T.~Sekihara and S.~Kumano,
  Phys.\ Rev.\ D {\bf 92}, no. 3, 034010 (2015).
  
\bibitem{aceti}
F.~Aceti, W.~H.~Liang, E.~Oset, J.~J.~Wu and B.~S.~Zou,
  Phys.\ Rev.\ D {\bf 86}, 114007 (2012).

\end{thebibliography}
\end{document}